\documentclass[twocolumn,10pt]{asme2e}

\usepackage{graphicx}
\usepackage{amsmath}
\usepackage{subfigure}
\usepackage{setspace}
\usepackage{multirow}
\usepackage{mathptmx}

\confshortname{GT2019}
\conffullname{ASME Turbo Expo 2019: Turbomachinery Technical Conference and Exposition}

\confdate{17-21}
\confmonth{June}
\confyear{2019}
\confcity{Phoenix, AZ}
\confcountry{USA}

\papernum{GT2019-90498}

\title{GENERALIZATION OF MACHINE-LEARNED TURBULENT HEAT FLUX MODELS APPLIED TO FILM COOLING FLOWS}

\author{Pedro M. Milani\thanks{Address all correspondence to this author.}
    \affiliation{
	Mechanical Engineering Department\\
	Stanford University\\
	Stanford, CA 94305\\
  	Email: pmmilani@stanford.edu
    }
}

\author{Julia Ling
    \affiliation{
    Director of Data Science\\
	Citrine Informatics\\
	Redwood City, CA 94063\\
    }	
}

\author{John K. Eaton
    \affiliation{
	Mechanical Engineering Department\\
	Stanford University\\
	Stanford, CA 94305\\
    }	
}

\begin{document}

\maketitle    

\begin{abstract}
{\it The design of film cooling systems relies heavily on Reynolds-Averaged Navier-Stokes (RANS) simulations, which solve for mean quantities and model all turbulent scales. Most turbulent heat flux models, which are based on isotropic diffusion with a fixed turbulent Prandtl number ($Pr_t$), fail to accurately predict heat transfer in film cooling flows. In the present work, machine learning models are trained to predict a non-uniform $Pr_t$ field, using various datasets as training sets. The ability of these models to generalize beyond the flows on which they were trained is explored. Furthermore, visualization techniques are employed to compare distinct datasets and to help explain the cross-validation results.}
\end{abstract}

\begin{nomenclature}
\entry{$\theta$}{Dimensionless temperature}
\entry{$u_i$}{Velocity component in the i-th direction}
\entry{$D$}{Hole diameter in film cooling flows}
\entry{$BR$}{Blowing ratio in a jet in crossflow configuration}
\entry{$d$}{Distance to the nearest wall}
\entry{$k$}{Turbulent kinetic energy}
\entry{$\epsilon$}{Turbulent dissipation rate}
\entry{$\nu_t$}{Eddy viscosity calculated by realizable $k-\epsilon$ model [$m^2/s$]}
\entry{$\alpha_t$}{Turbulent diffusivity [$m^2/s$]}
\entry{$Pr_t$}{Turbulent Prandtl number $\nu_t/\alpha_t$}
\entry{$V^{(i)}$}{Volume of the i-th computational cell}
\entry{GDH}{Gradient Diffusion Hypothesis}
\entry{ML}{Machine Learning}
\entry{RF}{Random Forest}
\entry{PCA}{Principal Component Analysis}
\entry{t-SNE}{t-Distributed Stochastic Neighbor Embedding}
\end{nomenclature}

\section {INTRODUCTION}

Gas turbine blades operate in extremely high temperature environments, and thus require well-engineered cooling techniques to meet desired lifespans. Discrete hole film cooling is one such technique \cite{bogard} in which coolant (usually air diverted from the compressor stage) is ejected from holes in the blade to create a protective layer of cooler fluid on the blade's outer surface. When designing film cooling schemes, engineers aim to minimize the coolant flow while providing enough thermal protection. Computational fluid dynamics simulations are therefore used extensively in the design process to predict the mean temperature field arising from a certain film cooling configuration. Scale-resolving simulations, such as Large Eddy Simulations (LES), are typically able to accurately predict the mean temperature field in film cooling flows. However, they cannot feasibly be used to design complex geometries due to their prohibitive computational cost. Therefore, Reynolds-Averaged Navier Stokes (RANS) simulations are the workhorse in the industry, and will remain so for the foreseeable future. The downside of using RANS simulations is that their results are typically inaccurate, particularly for the mean temperature field. Recent results from Nikiparto et al. \cite{nikparto_igtiblade} exemplify this - all their RANS simulations failed to match experimental data for adiabatic effectiveness in a turbine blade. The long-term goal of the present work is to address this insufficiency.

In this paper, the flow is considered incompressible and the temperature is assumed to behave as a passive scalar. The equation governing the mean temperature distribution $\bar{\theta}$ that is solved in a RANS simulation is given in Eq.~\ref{eq_raad}, where $u_i$ represents the velocity field and $\alpha$ is the molecular diffusivity. The unclosed term $\overline{u_i'\theta'}$ is the turbulent heat flux, for which a model needs to be prescribed. The most widely used closure is the gradient diffusion hypothesis (GDH), which assumes the heat flux is proportional to the mean temperature gradient as shown in Eq.~\ref{eq_gdh}. To specify the turbulent diffusivity $\alpha_t$, solvers usually employ a fixed turbulent Prandtl number, $Pr_t = \nu_t / \alpha_t$, where $\nu_t$ is the eddy viscosity field calculated by the momentum solver. The value chosen, usually $Pr_t = 0.85$, is backed out of experimental profiles in the log-layer of a flat plate turbulent boundary layer \cite{kays}. 

\begin{equation}
\frac{\partial}{\partial{x_i}}(\bar{u_i}\bar{\theta}) = \frac{\partial}{\partial{x_i}}(\alpha \frac{\partial{\bar{\theta}}}{\partial{x_i}}) - \frac{\partial}{\partial{x_i}} \overline{u_i'\theta'}
\label{eq_raad}
\end{equation}
\begin{equation}
\overline{u_i'\theta'} = -\alpha_t \frac{\partial{\bar{\theta}}}{\partial{x_i}}
\label{eq_gdh}
\end{equation}

Previous work has shown that the traditional model described earlier is inappropriate in film cooling. Kohli and Bogard \cite{kohli} assumed the GDH was valid and measured highly non-uniform values of $Pr_t$ in film cooling flows. Muppidi and Mahesh \cite{muppidi_dns} and Schreivogel et al. \cite{schreivogel} studied turbulent mixing in different jet in crossflow geometries and found localized regions of counter-gradient diffusion, contradicting the fixed $Pr_t$ and the isotropic diffusivity assumptions. Oliver et al. \cite{oliver2017implicit} found misalignment between the turbulent heat flux and the mean temperature gradient vectors, which also contradicts the GDH of Eq.~\ref{eq_gdh}.

A few other models for the turbulent heat flux $\overline{u_i'\theta'}$ have been proposed, including the generalized gradient diffusion hypothesis (GGDH) of Daly and Harlow \cite{daly} and the higher order generalized gradient diffusion hypothesis (HOGGDH) of Abe and Suga \cite{abe}. However, Ling et al. \cite{julia_analysisJICF} and Ryan et al. \cite{ryan_skewed} tested these higher order models in film cooling flows and saw modest improvement at best. They also obtained significantly better mean temperature distribution by using a turbulent diffusivity field calculated from LES data, showing that the prescription of $\alpha_t$ in Eq.~\ref{eq_gdh} is an important source of error. So, due to the poor performance of the traditional models and lack of clearly superior alternatives, the sole focus of the present work will be improving turbulent heat flux modeling. For that, a machine learning approach will be employed.

Machine learning is a collection of algorithms that process large amounts of data and attempt to extract patterns from that data \cite{MLbook}. In the past 10 years, it has attracted considerable attention due to extraordinary results obtained in fields like image recognition and machine translation. Its use in the turbulence modeling community is incipient, but promising. Ling et al. \cite{julia_deepnn} proposed a deep neural network architecture that respects Galilean invariance to predict the turbulent anisotropy tensor, and demonstrated improved results compared to linear eddy viscosity models. Sandberg et al. \cite{sandberg_igti2018} used gene expression programming to obtain non-linear closed-form expressions for the anisotropy tensor and the turbulent diffusivity in 2D slot film cooling geometries, which significantly improved RANS predictions. Singh et al. \cite{singh_machinelearning2017} employed field inversion (through adjoint optimization) combined with neural networks to improve closure equations for the momentum solver, which improves lift predictions in airfoils with separated flows. The present paper is based on the machine learning framework first proposed by Milani et al. \cite{milani_approach2017}, but with key differences that will be highlighted in Section 2. In their work, a random forest was used to predict the turbulent diffusivity in film cooling flows, and it produced significant improvements in mean temperature predictions, particularly close to the wall.

Robustness of such machine-learned models is an outstanding topic. In particular, how closely related must the training data be to a flow configuration of interest for the model to produce improved results? Ultimately, this question is crucial because the usefulness of any turbulence model based on machine learning techniques depends on it. The work reported in this paper investigates this question in the context of turbulent heat flux models for film cooling applications. Section 2 explains the model and describes the datasets and experiments used to address generalization. Section 3 shows results for the turbulent Prandtl number obtained in inclined jets in crossflow with differently trained models. Section 4 presents the mean temperature obtained by solving Eq.~\ref{eq_raad} using the $Pr_t$ fields predicted in Section 3. Section 5 introduces visualization techniques and employs them in an attempt to explain the results of Sections 3 and 4. Finally, Section 6 presents conclusions and ideas for future work.

\section {METHODOLOGY}

\subsection{Machine Learning Model}
The machine learning model for turbulent heat flux will be discussed in this subsection. In summary, the GDH of Eq.~\ref{eq_gdh} is employed, but the turbulent diffusivity is not prescribed with a fixed turbulent Prandtl number. Instead, a random forest (RF) algorithm is used to predict $Pr_t$ at each point in the domain based on local mean flow information. As a supervised learning algorithm, the RF must be trained in datasets where the mean flow information is known together with the correct value of $Pr_t$.

To train the algorithm, well-validated, high-fidelity simulations are needed. The local mean quantities that are assumed to govern the value of the turbulent Prandtl number are extracted from the LES: they are the mean velocity gradient, $\nabla \bar{ \mathbf{u}}$, the mean temperature gradient, $\nabla \bar{ \theta}$, the eddy viscosity ratio, $\nu_t / \nu$, and the Reynolds number based on distance to the nearest wall $Re_d = \sqrt{k}d/\nu$. The velocity and temperature gradients are non-dimensionalized using local turbulent scales built with $k$ and $\epsilon$. Also, to enforce Galilean invariance of the output diffusivity as recommended by Ling et al. \cite{julia_invariance}, an invariant basis is contructed based on the gradients and this basis is used as the input to the ML algorithm. A complete list with the 19 inputs can be found in Milani et al. \cite{milani_approach2017}. It is important to note that the high-fidelity simulation directly provides mean velocity and temperature; to obtain the turbulence quantities such as $k$, $\epsilon$, and $\nu_t$, the mean velocity field is frozen and RANS turbulence equations are solved on the LES mesh (similar to the approach of Sandberg et al. \cite{sandberg_igti2018}). In the present case, realizable $k-\epsilon$ equations of Shih et al. \cite{shih_realizable} are solved using ANSYS Fluent 18.0. Also note that this approach differs from the one presented in Milani et al. \cite{milani_approach2017}: in their work, RANS variables were used as inputs. In the present work, LES mean quantities (with a $k-\epsilon$ solution for turbulence) are used instead. This makes the current model closer to a ``true'' turbulence model since it takes in the high-fidelity mean quantities and tries to map them to a turbulence quantity of interest. When the model is trained on RANS quantities, it has to account not only for turbulence, but also for errors incurred by the original RANS simulation (since the RANS mean velocity and temperature fields might incorrectly portray reality).

To calibrate the random forest, the ``true'' value of $Pr_t$ in each cell is also needed. To obtain that, the local value of $\alpha_t$ is backed out of the high fidelity simulation: information on the turbulent heat flux $\overline{u_i'\theta'}$ is employed to algebraically extract $\alpha_{t,LES}$ at each cell of the domain. In Milani et al. \cite{milani_approach2017}, a simple least squares solution for $\alpha_{t,LES}$ is used. The present work uses a slight modification of this approach, which mitigates the underestimation of turbulent diffusion observed previously. A weighted average of the least squares sum is used, where the weight $F$ in each of the three directions decreases as the mean advection in that direction increases. The reasoning behind this is that in a direction with strong mean advection (generally the streamwise direction), the contribution of turbulent mixing to the overall mean scalar transport is negligible regardless of the value of $\alpha_t$. Thus the information from that direction should not influence the extracted $\alpha_{t,LES}$. In contrast, it is more important to predict the turbulent transport accurately in directions in which the mean advection is weak (for example, the spanwise direction in most flows). Empirically, this generated anywhere between no difference and significant improvement across the flow geometries in this paper when compared to the approach of Milani et al.\cite{milani_approach2017}. Equations~\ref{eq_alphat} and \ref{eq_F} summarize how $\alpha_{t,LES}$ is extracted from LES quantities. The turbulent Prandtl number used to train the RF is calculated as $Pr_t = \nu_t / \alpha_{t,LES}$, and is clipped between $0.01$ and $100$.

\begin{equation}
\alpha_{t,LES} = - \frac{\sum_{i=1}^3 \overline{u_i'\theta'} \frac{\partial{\bar{\theta}}}{\partial{x_i}} F_i} {\sum_{j=1}^3 \frac{\partial{\bar{\theta}}}{\partial{x_j}} \frac{\partial{\bar{\theta}}}{\partial{x_j}} F_j}
\label{eq_alphat}
\end{equation}
\begin{equation}
F_i \equiv \frac{1}{\sqrt{\overline{u_j'\theta'} \ \overline{u_j'\theta'}} + \bar{u}_i \bar{\theta}}
\label{eq_F}
\end{equation}

The random forest algorithm is used to map from the 19 inputs to $ln(Pr_t)$ in each computational cell of the high-fidelity simulations in the training set. The natural logarithm is important to ensure that the relative, not absolute, error in $Pr_t$ is minimized (e.g., if the true $Pr_t$ is 0.1 and the prediction is 0.2, the model is penalized just as much as if the true $Pr_t$ were 10 and the predicition were 20). The RF is chosen because Ling and Templeton \cite{julia_algorithms} showed that it achieves good results in turbulence modeling, and unlike neural networks it is robust to noise and to outliers and is also mostly insensitive to the hyperparameters. In this paper, 500 trees were used and criteria were set for early stopping (maximum depth of 25 and minimum number of samples required to split a node of the tree of 0.01\% of the total training set size). This set of hyperparameters leads the RF to perform similarly to the one of Milani et al. \cite{milani_approach2017}, but creates more concise models and speeds up training on large datasets. The model was implemented in Python using the scikit-learn library \cite{pedregosa}. For more information on random forests, consult Louppe's comprehensive review \cite{louppe}.

\subsection{Datasets}

\begin{figure}[h]
\begin{center}
\includegraphics [width = 80mm]{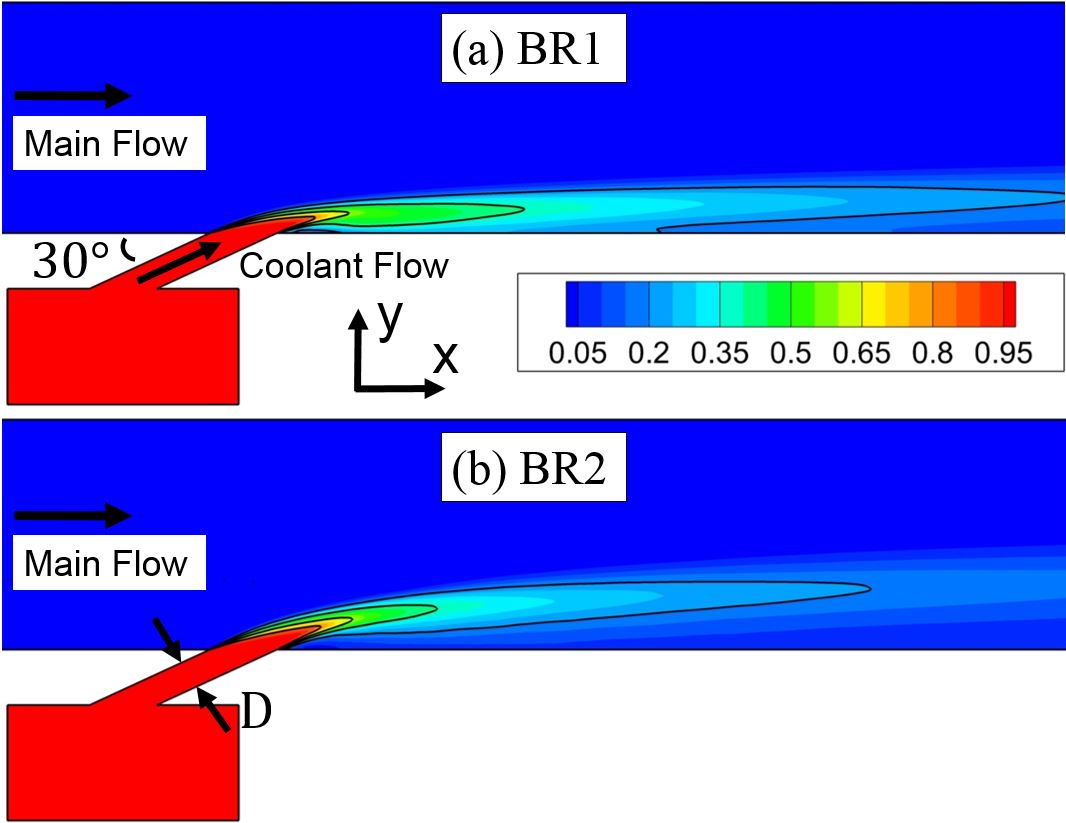}
\end{center}
\caption{Center spanwise plane showing contours of $\bar{\theta}$ in the baseline geometry for two distinct blowing ratios, (a) $BR=1$ and (b) $BR=2$.}
\label{figure-br1_br2_show}
\end{figure}

\begin{figure}[h]
\begin{center}
\includegraphics [width = 85mm]{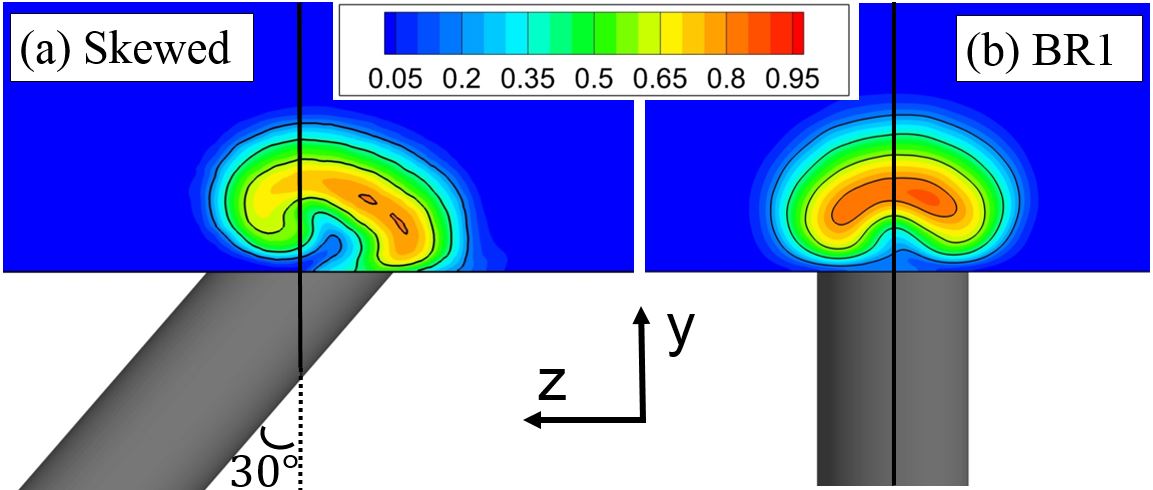}
\end{center}
\caption{Streamwise planes located at $X/D=2$ showing contours of $\bar{\theta}$ in the (a) \textit{Skewed} and (b) \textit{BR1} cases. The hole is shown in gray and a vertical line at $Z/D=0$ indicates the center of the hole as it meets the bottom wall. The main flow direction is out of the page.} 
\label{figure-skewed_show}
\end{figure}

\begin{figure}[h]
\begin{center}
\includegraphics [width = 70mm]{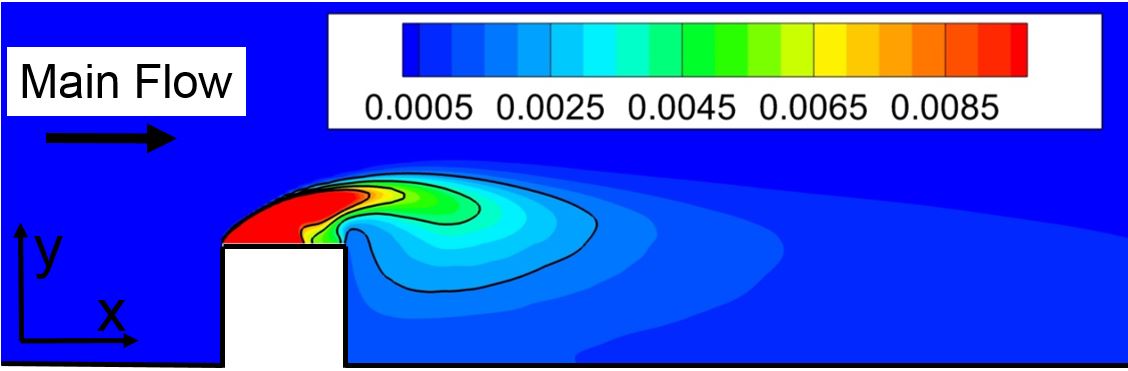}
\end{center}
\caption{Center spanwise plane of \textit{Cube} case, showing contours of mean temperature $\bar{\theta}$.} 
\label{figure-cube_show} 
\end{figure}

Four datasets representing different configurations are used in the present work. They are: 
\begin{enumerate}
  \item \textit{BR1} (shown in Fig.~\ref{figure-br1_br2_show}(a) and Fig.~\ref{figure-skewed_show}(b))
  \item \textit{BR2} (shown in Fig.~\ref{figure-br1_br2_show}(b))
  \item \textit{Skewed} (shown in Fig.~\ref{figure-skewed_show}(a))
  \item \textit{Cube} (shown in Fig.~\ref{figure-cube_show})
\end{enumerate} In all the cases, a high-quality LES or direct numerical simulation (DNS) is available, where both momentum and passive scalar equations are solved (the latter as a proxy to temperature) and statistics on turbulent heat flux $\overline{u_i'\theta'}$ were collected. The four simulations show good agreement against experimental data. The rest of this subsection describes them in-depth.

The first two cases, \textit{BR1} and \textit{BR2}, are run in a baseline film cooling geometry, which contains a single circular cooling hole of diameter $D$ discharging into a square channel of side $8.6D$. The hole is inclined $30^\circ$ and its length is $L/D = 4.1$. The fully turbulent incoming boundary layer has thickness $\delta / D = 1.5$ measured $2D$ upstream of the hole. Note that, since the simulation is incompressible, the density ratio is exactly 1.0. The hole is fed from a plenum underneath the channel, where the dimensionless temperature is $\bar{\theta}=1$; at the channel inlet, the temperature is set to $\bar{\theta}=0$. All walls are adiabatic and the molecular Prandtl number is $Pr=1.0$. The two cases are distinct due to their blowing ratio, $BR = \rho U_{j} / \rho U_{c}$, where $U_{j}$ is the bulk velocity in the jet and $U_{c}$ is the bulk velocity in the main channel. The first has $BR=1$ (and $Re_D = 2,900$) and the second has $BR=2$ (and $Re_D = 5,800$), where $Re_D = U_j D / \nu$ is the Reynolds number based on the hole diameter and jet bulk velocity. The two LES's followed the methodology of Bodart et al. \cite{julien}, and were run using the incompressible solver Vida from Cascade Technologies. The meshes ($40.1M$ cells and $48.3M$ cells respectively) are wall-resolved in the bottom wall and cooling hole ($y^+ < 1.5$). A posteriori analysis showed that the mean subgrid scale viscosity (which uses the Vreman model \cite{vreman_sgsmodel}) is negligible compared to the laminar viscosity. The mean velocity and temperature compare satisfactorily against 3D MRI data obtained in the same geometry. More simulation details and the validation are presented in Milani et al. \cite{milani2019enriching}. Figure~\ref{figure-br1_br2_show} shows the centerplane from those two simulations.

The third dataset is the \textit{Skewed} case, produced by Folkersma et al. \cite{mikko}. It is the same as the baseline geometry, except that the cooling hole has a compound angle of injection: inclined $30^\circ$ about the streamwise and spanwise directions (see Fig.~\ref{figure-skewed_show}(a)). The channel has a rectangular cross-section ($17.2D$ by $8.6D$) and the incoming boundary layer has thickness $\delta / D = 1.9$ measured $2D$ upstream of the hole. The density and blowing ratios are both 1.0, and $Re_D = 5,800$. The walls are adiabatic and the molecular Prandtl number is $Pr=1.0$. The LES has around $101M$ cells and is also wall-resolved. It is described in detail and validated against MRI experiments in Folkersma et al. \cite{mikko}.

\vspace{-1em}

The last dataset is the \textit{Cube}, computed by Rossi et al. \cite{rossi}. In this flow, a turbulent boundary layer meets a wall-mounted cube of side $D$, which causes flow separation behind it. A circular source centered on the top surface of the cube injects small amounts of heated fluid, which creates a non-trivial mean temperature distribution (see Fig.~\ref{figure-cube_show}). The Reynolds number based on cube height and free-stream velocity is $Re_D=5000$ and the molecular Prandtl number is set to $Pr=0.285$. This dataset, despite not being directly relevant to film cooling, was chosen because of the availability of a high-fidelity simulation with a scalar contaminant. Furthermore, it contains features relevant to film cooling such as turbulent mixing within a separation bubble. A DNS was run in this case, and the results were validated against experimental data. Further details are available in Rossi et al. \cite{rossi}.

\subsection{Cross-Validation}

The current study leverages the concept of cross-validation. Cross-validation is an important tool in machine learning to evaluate the capacity of a model to generalize beyond its training data. It consists of training a model and testing its performance in different subsets of the total set of examples available. In the present paper, 4 different datasets are available. In total, 6 different random forests were trained to predict $Pr_t$ given local mean flow features. Those models were then applied to the two datasets of the baseline geometry, \textit{BR1} and \textit{BR2}. The performance of the models is dictated by how accurately the mean temperature field is predicted compared to the LES mean field. Table~\ref{table-rfs} summarizes the 6 trained models used.

Note that models are only tested on the \textit{BR1} and \textit{BR2} cases for the sake of brevity; the two cases were chosen because they illustrate physics of turbulent mixing in film cooling. The first 4 models were trained on each of the 4 datasets individually in order to understand how much physics each flow by itself can provide to the RF. The last two training sets contain all datasets except \textit{BR2} (which is applied to the \textit{BR2} case) and all datasets except \textit{BR1} (which is applied to the \textit{BR1} case) respectively. This tests the ability of models trained on several distinct datasets to generalize to an unseen case. 

In all cases, the algorithm is trained only on computational cells where the temperature gradient non-dimensionalized by the local turbulent length scale is above a given threshold. In this paper, the threshold used is $10^{-3}$, but the results are mostly insensitive to this parameter within a few orders of magnitude. This is done because the turbulent diffusivity $\alpha_{t,LES}$ should only be extracted according to Eq.~\ref{eq_alphat} in locations where the temperature gradient is not negligible; otherwise, the denominator approaches zero and the results become extremely noisy. At test time, predictions are only made in locations that obey the $10^{-3}$ cutoff; elsewhere, a default value of $Pr_t=0.85$ is employed.

\begin{table}[h]
\centering
\caption{Summary of the random forests trained.}
\label{table-rfs}
\renewcommand{\arraystretch}{0.9}
\begin{tabular}{|p{2cm} | p{5cm} |}
\hline
\textbf{Model} & \textbf{Trained on}\\
\hline
\verb|RF_b1| & Baseline geometry with BR=1\\
\verb|RF_b2| & Baseline geometry with BR=2\\
\verb|RF_s| & Skewed\\
\verb|RF_c| & Cube\\
\verb|RF_csb1| & Cube, Skewed, and BR1 cases\\
\verb|RF_csb2| & Cube, Skewed, and BR2 cases\\
\hline
\end{tabular}
\end{table}
\vspace{-1.75em}

\section {TURBULENT PRANDTL NUMBER RESULTS}

\begin{figure*}[h]
\begin{center}
\includegraphics [width = 152mm]{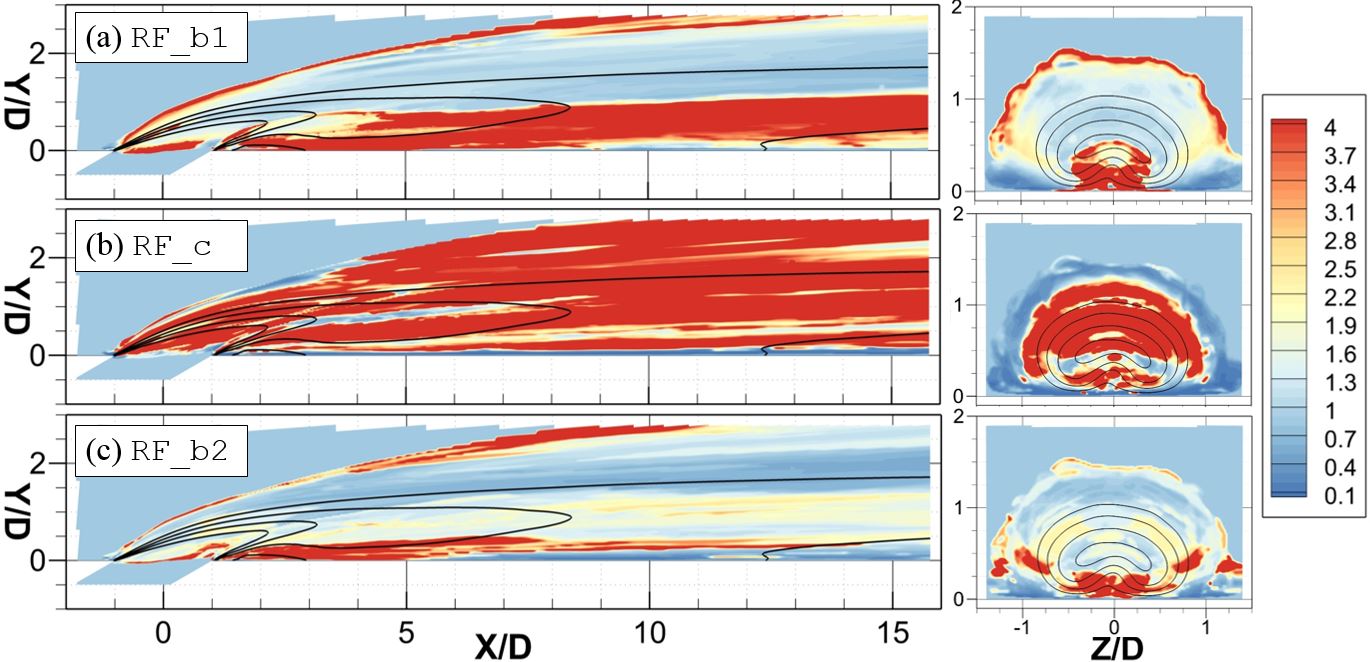}
\end{center}
\caption{Turbulent Prandtl number predicted by different models on the \textit{BR1} case. The plots also contain isocontour lines of mean temperature calculated by the LES at $\bar{\theta}=0.2, 0.4, 0.6, 0.8$. The plots on the left show center spanwise planes at $Z/D=0$ and plots on the right show streamwise planes at $X/D=2$. From top to bottom, the predictions of $Pr_t$ are made by (a) \texttt{RF\symbol{95}b1}, (b) \texttt{RF\symbol{95}c}, and (c) \texttt{RF\symbol{95}b2}.} 
\label{figure-br1_prt} 
\end{figure*}

The first step of the cross-validation study is to use the different models to predict the turbulent Prandtl number. Fig.~\ref{figure-br1_prt} shows contour plots with $Pr_t$ values predicted by three different random forests on the \textit{BR1} case. The plots on Fig.~\ref{figure-br1_prt}(a) have $Pr_t$ predictions from the model trained on the case \textit{BR1} itself. This is not realistic for practical purposes, because a machine learning model is expected to be employed for cases distinct from the ones it was trained on. However, it is useful in the context of this paper because it illustrates a performance upper bound: the best possible predictions can be expected when the ML model is trained on the same data on which it is applied. Note that this field is extremely similar to the exact $Pr_t$ field, extracted from the LES results using Eq.~\ref{eq_alphat} (not shown in Fig.~\ref{figure-br1_prt}). Some features of this ideal $Pr_t$ field become clear. First, it is highly non-uniform as previously reported (e.g. \cite{kohli}), so a fixed $Pr_t=0.85$ is likely inappropriate for this flow. The centerplane plot suggests that the bottom half of the jet requires high values of turbulent Prandtl number ($Pr_t > 4$), while in the top half moderate values around $0.85$ are more appropriate.

\begin{figure*}[h]
\begin{center}
\includegraphics [width = 152mm]{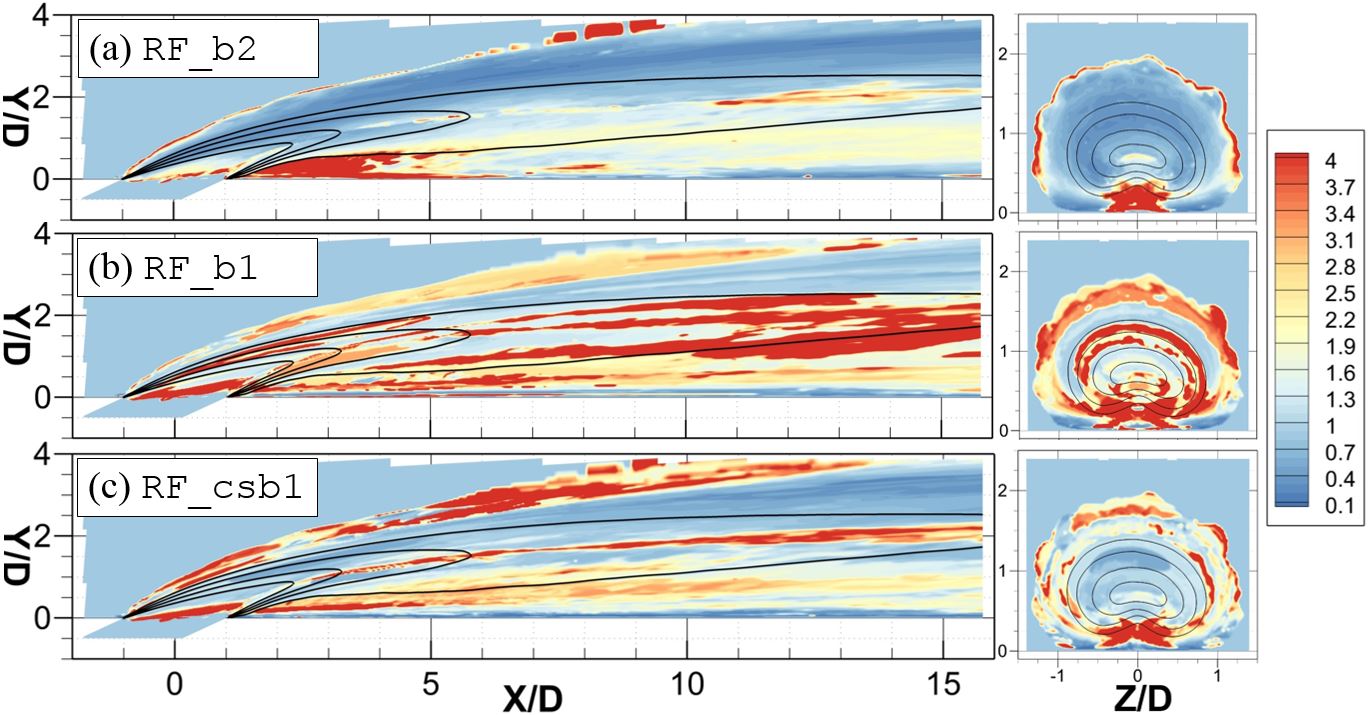}
\end{center}
\caption{Turbulent Prandtl number predicted by different models on the \textit{BR2} dataset. The plots are superimposed with isocontour lines of mean temperature from the LES at $\bar{\theta}=0.2, 0.4, 0.6, 0.8$. The plots on the left show center spanwise planes at $Z/D=0$ and plots on the right show streamwise planes at $X/D=2$. From top to bottom, the predictions of $Pr_t$ are made by (a) \texttt{RF\symbol{95}b2}, (b) \texttt{RF\symbol{95}b1}, and (c) \texttt{RF\symbol{95}csb1}.} 
\label{figure-br2_prt}
\end{figure*}

Figure~\ref{figure-br1_prt}(b) shows the $Pr_t$ field predicted by the model that is trained only on the \textit{Cube} case. It produces consistently high values of $Pr_t$ throughout the whole domain, contradicting the predictions of the \verb|RF_b1| model. This effectively translates to low turbulent diffusivities almost everywhere in Eq.~\ref{eq_raad}. The poor predictions are not surprising, since the \textit{Cube} flow is significantly different from the \textit{BR1} flow and therefore might not, by itself, provide a good training set. Finally, Fig.~\ref{figure-br1_prt}(c) illustrates the turbulent Prandtl number predicted by a model trained only on the \textit{BR2} case. The training and evaluation sets have the same geometry, but the blowing ratio is a crucial parameter in film cooling, and doubling it from 1 to 2 significantly alters the resulting mean flow and temperature. So, this scenario tests the ability of the model to generalize beyond its training set. Despite some differences, the field in Fig.~\ref{figure-br1_prt}(c) has notable similarities to the one produced by \verb|RF_b1|. For instance, high values of $Pr_t$ in the bottom part of the jet, particularly right after injection, and moderate values of $Pr_t$ in the top shear layer.

Figure~\ref{figure-br2_prt} shows turbulent Prandtl number fields predicted in the \textit{BR2} case with three different random forests. Figure~\ref{figure-br2_prt}(a) is generated by \verb|RF_b2|, which again shows a performance upper bound for this ML framework. Again, higher values or $Pr_t$ are present on the bottom half of the jet, which is particularly evident right after injection (as shown in the streamwise plot). Figure~\ref{figure-br2_prt}(b) contains predictions from \verb|RF_b1|, which seem to overestimate $Pr_t$ in most of the domain, particularly towards the top and sides of the jet. Figure~\ref{figure-br2_prt}(c), was produced with a model trained on all datasets except for \textit{BR2}, namely \verb|RF_csb1|. It is far from ideal, but it shows some improvement over the field produced by \verb|RF_b1| because the overestimation of $Pr_t$ is not as severe. Furthermore, in Fig.~\ref{figure-br2_prt}(c) the trend of higher $Pr_t$ in the bottom half of the jet and moderate $Pr_t$ in the top half of the jet becomes apparent. 

Some of the results in this section are consistent with previous work in the literature. For example, Ling et al. \cite{julia_wallmodel} argued that the asymptotic behavior of $\overline{v'\theta'}$ and $\overline{v'u'}$ is distinct as you approach an adiabtic wall, so a fixed ratio of $\nu_t$ to $\alpha_t$ is invalid. That led them to propose a near-wall correction in which $Pr_t$ would exponentially decrease at $y^+ < 70$, which produced improved results in slot film cooling geometries \cite{julia_wallmodel}. The predictions in both \textit{BR1} and \textit{BR2} cases show that the random forests have learned this pattern: a consistent thin blue strip right above the bottom wall shows that most of the ML models predict reduced values of $Pr_t$ when $y^+$ is small. The high values of turbulent Prandtl number found right after injection and close to the bottom wall in the two datasets are also consistent with the findings of Milani and Eaton \cite{milani_australasian2018}. They used MRI data and an optimization approach to conclude that this region should have extremely low values of vertical turbulent diffusivity across five distinct film cooling flows.

\section {MEAN TEMPERATURE RESULTS}

The previous section compared different models to predict the turbulent Prandtl number. However, the ultimate goal of this work is to improve mean temperature predictions. Thus, the preferred way to evaluate different $Pr_t$ fields is to solve Eq.~\ref{eq_raad} using them and then compare the resulting mean temperature against the LES mean temperature. Eq.~\ref{eq_raad} is solved on the LES mesh with the LES velocity field. The correct velocity field is used instead of a RANS velocity field to isolate the errors in $\bar{\theta}$ due to the turbulent heat flux model. This contrasts with our previous work \cite{milani_approach2017, milani_interpretation2018} where the temperature field was calculated using the RANS velocity field. It is important to note that the ML models predict the $Pr_t$ field once based on the LES flow information, including the LES mean concentration gradient $\nabla \bar{ \theta}$, and then this $Pr_t$ field is used to solve the temperature transport equation. While this method is acceptable for studying model generality, it is not consistent if used directly on a RANS simulation. In that case, the RF model would need to be built into the RANS solver and used iteratively, since the predicted $Pr_t$ field is a function of the underlying true temperature gradient.

\subsection{Baseline geometry with $BR=1$}

\begin{figure}[h]
\begin{center}
\includegraphics [width = 90mm]{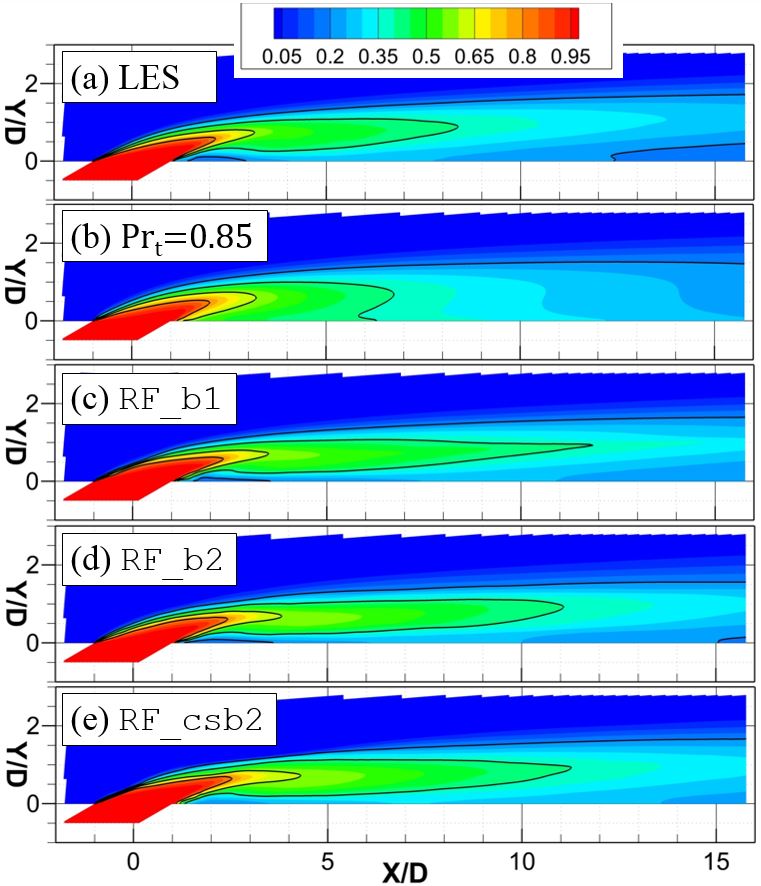}
\end{center}
\caption{Contour plots of mean temperature $\bar{\theta}$ in the center spanwise plane of the \textit{BR1} case. (a) shows the LES results, (b) shows the results of using a uniform Prandtl number, set at $Pr_t=0.85$, and (c)-(e) show the results of using different ML fields for $Pr_t$. Isocontour lines of mean temperature are superimposed, at $\bar{\theta}=0.2, 0.4, 0.6, 0.8$.} 
\label{figure-br1_spanwise} 
\end{figure}

Figure~\ref{figure-br1_spanwise} shows contours of $\bar{\theta}$ obtained with selected models in the \textit{BR1} case. Figure~\ref{figure-br1_spanwise}(a) contains the mean temperature field from the LES, which was validated against experimental data and is thus considered the ``ground truth'' against which to compare results from different models. Fig.~\ref{figure-br1_spanwise}(b) shows the temperature resulting from the traditional fixed turbulent Prandtl number assumption, $Pr_t=0.85$. As evidenced by the results, the turbulent mixing seems to be overestimated in most of the domain, particularly close to the wall, since the high scalar values in the jet core tend to diffuse towards the bottom wall.

Figure~\ref{figure-br1_spanwise}(c) contains the temperature predictions of the random forest trained exclusively on the \textit{BR1} case (which produced the $Pr_t$ field shown in Fig.~\ref{figure-br1_prt}(a)), and therefore represents the results that this framework yields with a perfectly trained ML algorithm. As seen in Fig.~\ref{figure-br1_prt}(a), there are high values of $Pr_t$ in the bottom half of the jet which imply lower values of $\alpha_t$ there. This fixes the problem of too much mixing in this region and generates better temperature results close to the wall. However, this model is still deficient: the $\bar{\theta}=0.4$ isocontour shows that in the core of the jet the model fails to capture the correct turbulent heat flux. This points the finger at the model form (the GDH of Eq.~\ref{eq_gdh}) since, even with a perfect mean velocity and $Pr_t$ distribution, the mean temperature is still incorrectly calculated.

Figures~\ref{figure-br1_spanwise}(d) and (e) show results that are qualitatively similar to Fig.~\ref{figure-br1_spanwise}(c). This suggests that a random forest trained on the \textit{BR2} case captures the turbulent heat flux approximately as well as one that is perfectly trained. Furthermore, even though the \textit{Cube} is a poor training set for this case (\verb|RF_c| predicts diffusivities that are too low almost everywhere and, consequently, a poor temperature field), the random forest trained with \textit{Cube}, \textit{Skewed}, and \textit{BR2} datasets (\verb|RF_csb2| shown in Fig.~\ref{figure-br1_spanwise}(e)) performs fairly well on this problem.

\begin{figure}[h]
\begin{center}
\includegraphics [width = 90mm]{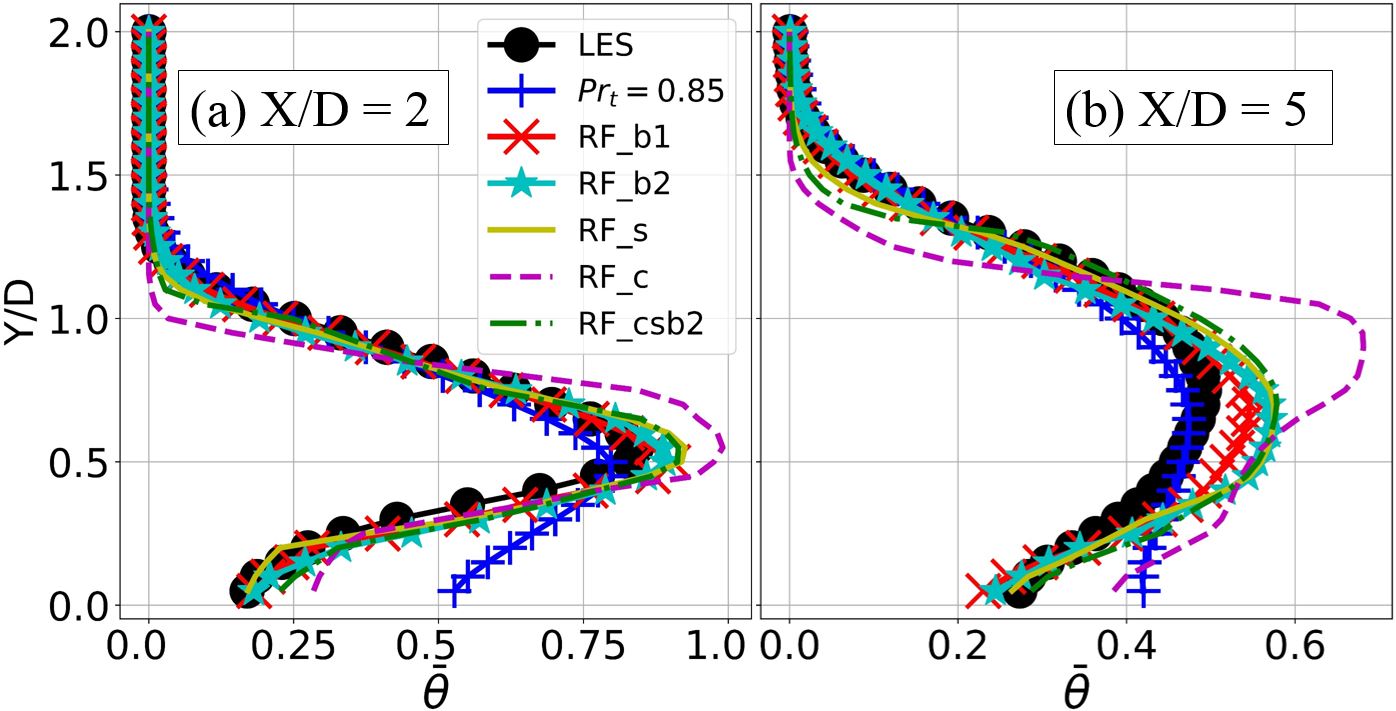}
\end{center}
\caption{Plots showing $\bar{\theta}$ vs $Y/D$ in the \textit{BR1} case at $Z/D=0$ and two different streamwise locations: (a) at $X/D=2$ and (b) at $X/D=5$.} 
\label{figure-br1_profiles} 
\end{figure}

Line plots provide more quantitative comparisons between distinct models and the LES results. Figure~\ref{figure-br1_profiles} shows vertical profiles of mean temperature in the \textit{BR1} case on the center spanwise plane, at two different streamwise locations. The black lines with circles show the temperature results of the LES, which is what the models are aiming to replicate. The problem with a fixed $Pr_t$ is again evident: too high a diffusivity smooths the profile, overestimates values close to the wall, and shifts the temperature peak down. As expected, the predictions from \verb|RF_b1| match the LES data more closely, particularly towards the wall. However, the peak temperatures are now overpredicted, potentially due to underpredicted lateral spreading. In these plots, it is possible to see that the other models predict profiles similar to the one predicted by \verb|RF_b1|, confirming their ability to generalize to the \textit{BR1} dataset. The exception is the model trained only on the \textit{Cube} (\verb|RF_c|), which predicts very poor mean temperature profiles due to uniformly high $Pr_t$. This shows that training exclusively on the \textit{Cube} dataset is not enough to predict acceptable results in the \textit{BR1} dataset.
 
\begin{figure}[h]
\begin{center}
\includegraphics [width = 54mm]{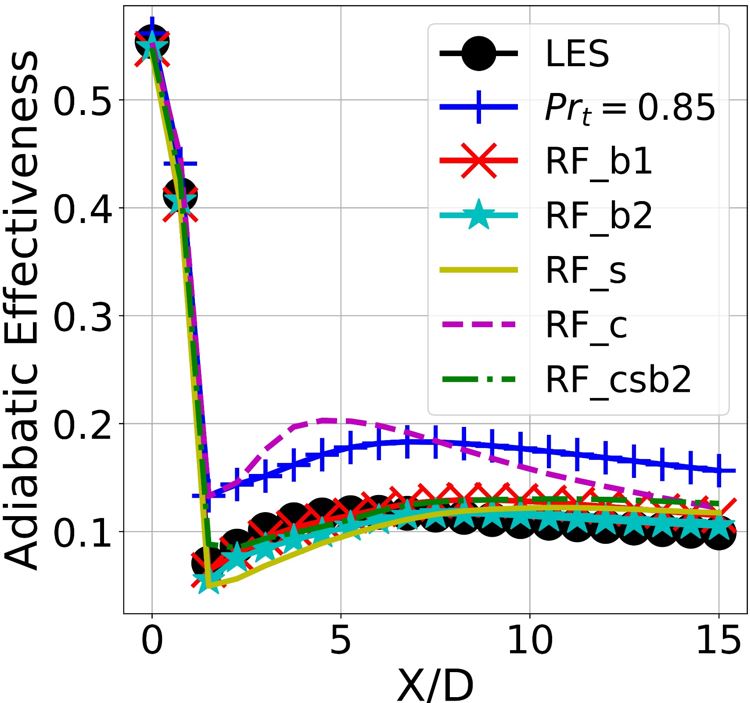}
\end{center}
\caption{Spanwise-averaged adiabatic effectiveness in the \textit{BR1} case. Averaging is performed between $Z/D=-1$ and $Z/D=1$.} 
\label{figure-br1_adiaeff} 
\end{figure}

The adiabatic effectiveness is a crucial quantity of interest in the film cooling literature, because it determines the ability of a certain configuration to shield the bottom wall from the high temperatures in the freestream. In the current problem setup, the adiabatic effectiveness is simply $\bar{\theta}$ evaluated at the wall ($Y/D=0$), since the simulations use adiabatic boundary conditions. Figure~\ref{figure-br1_adiaeff} shows a comparison of the spanwise-averaged adiabatic effectiveness for all models. As suggested before, the fixed $Pr_t$ model significantly overpredicts this quantity due to high turbulent heat flux in the vertical direction. The machine learning models, with the exception of \verb|RF_c|, perform well. The adiabatic effectiveness they predict tracks the LES curve closely, including the sharp dip right after injection, due to the improved values of $Pr_t$ close to the bottom wall.

\subsection{Baseline geometry with $BR=2$}

\begin{figure}[h]
\begin{center}
\includegraphics [width = 90mm]{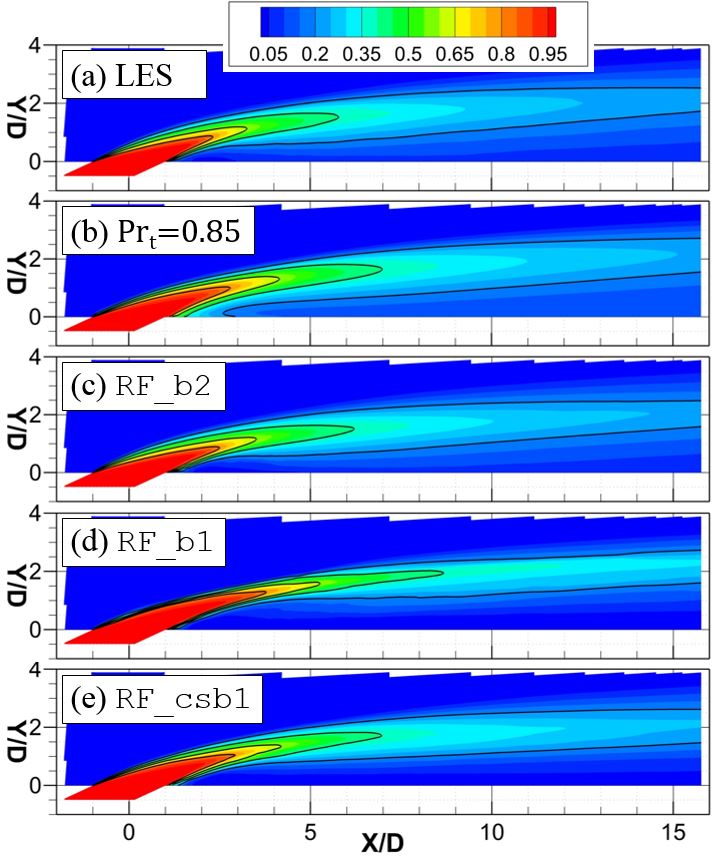}
\end{center}
\caption{Contour plots of mean temperature $\bar{\theta}$ in the center spanwise plane of the \textit{BR2} case. (a) shows the LES results, (b) shows the results of using $Pr_t=0.85$ everywhere, and (c)-(e) show the results of using different ML models for $Pr_t$. Isocontour lines of mean temperature are superimposed, at $\bar{\theta}=0.2, 0.4, 0.6, 0.8$} 
\label{figure-br2_spanwise} 
\end{figure}

\begin{figure}[h]
\begin{center}
\includegraphics [width = 90mm]{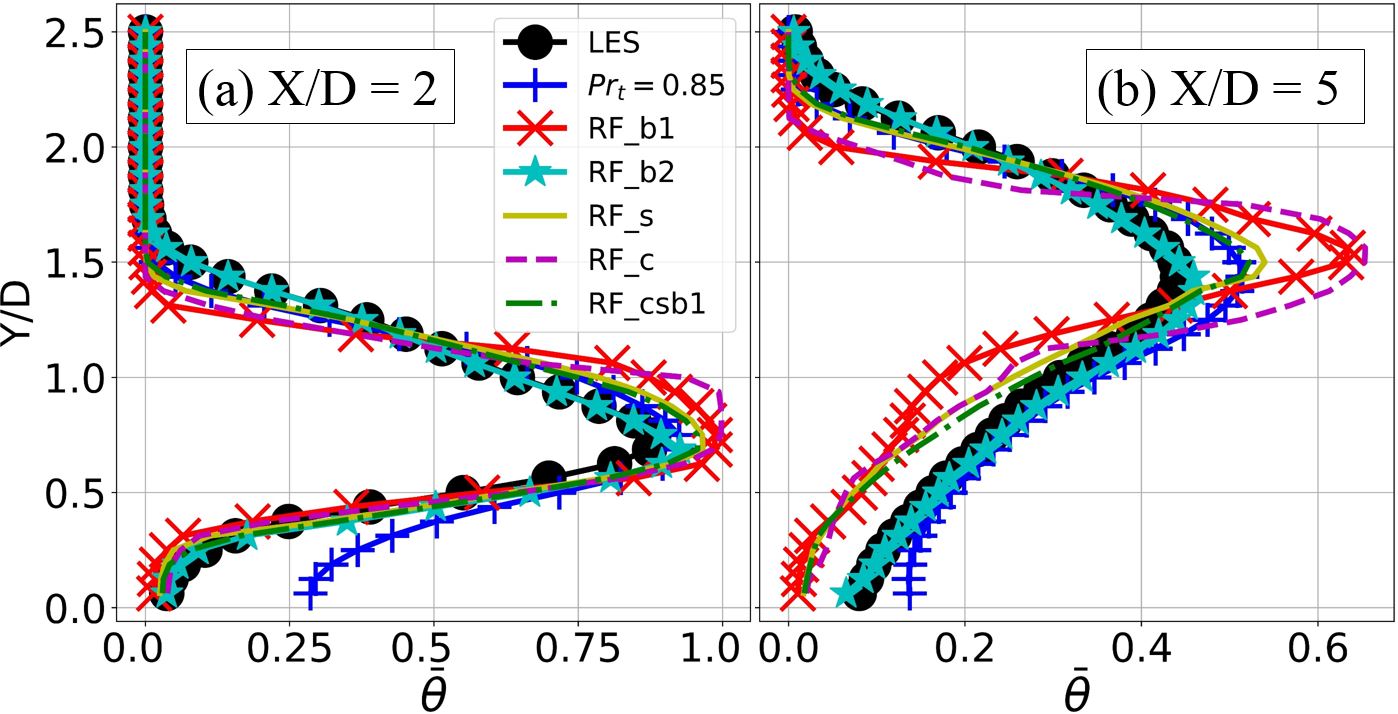}
\end{center}
\caption{Profile of $\bar{\theta}$ vs $Y/D$ in the \textit{BR2} case at $Z/D=0$ and two different streamwise locations: (a) at $X/D=2$ and (b) at $X/D=5$.}
\label{figure-br2_profiles} 
\end{figure}

A similar analysis is performed on the second test set, in which the goal is to predict the mean temperature field in the \textit{BR2} case. Figure~\ref{figure-br2_spanwise} contains contour plots of mean temperature in the centerplane of the \textit{BR2} geometry and Fig.~\ref{figure-br2_profiles} contains vertical profiles of mean temperature. Figure~\ref{figure-br2_spanwise}(a) shows LES results, which again serve as the ``ground truth'' that the distinct turbulent heat flux models are trying to replicate. The difference between this flow and \textit{BR1} is evident: the jet penetrates further into the main stream due to its higher momentum, and almost no scalar makes it all the way down to the bottom wall. The result from the typical $Pr_t=0.85$ model is shown in Fig.~\ref{figure-br2_spanwise}(b). Errors are small in most of the domain; the exception is close to the wall, right after injection. In the LES, very little mixing occurs there, but the fixed turbulent Prandtl number calculation predicts the coolant diffusing into that region.

\vspace*{-0.5cm}

When a perfectly trained ML model is applied, and the scalar equation is solved with the $Pr_t$ field shown in Fig.~\ref{figure-br2_prt}(a), the mean temperature predictions are excellent, as shown in Fig.~\ref{figure-br2_spanwise}(c) and in the profiles of Fig.~\ref{figure-br2_profiles}. This suggests that in this particular case, unlike in the \textit{BR1} case shown before, the model form of Eq.~\ref{eq_gdh} is appropriate and all that is needed to model the turbulent heat flux well is to spatially adjust the value of $Pr_t$. One potential reason is the importance of anistropy in the turbulent transport. In the \textit{BR1} case, the jet stays closer to the wall, which acts to damp vertical eddies making turbulence more anisotropic. When the jet is farther from the wall, as in the \textit{BR2} case, spanwise and wall-normal transport are probably more similar, so the isotropic formulation of the GDH becomes a better approximation. 

Interestingly, a machine learning model that is only trained on the \textit{BR1} case performs poorly on the present case, as shown in Fig.~\ref{figure-br2_spanwise}(d) and in Fig.~\ref{figure-br2_profiles}. The Prandtl number predicted is generally too high, which leads to low diffusivity values and incorrect mean temperature distributions. When the machine learning algorithm is trained on all cases except for \textit{BR2} (i.e., the \verb|RF_csb1| model), the $\bar{\theta}$ field significantly improves compared to the one produced by the \verb|RF_b1| model. Most of the improvement comes from adding the \textit{Skewed} case to the training set, since individually both the \textit{BR1} and the \textit{Cube} cases are not sufficient to produce good results. However, the predictions of the \verb|RF_csb1| model are still noticeably worse than the temperature predicted by the perfectly trained model. This suggests that generalization is the bottleneck to obtain good results with an ML model in the \textit{BR2} dataset. It is not sufficient to train the algorithm on the three other cases (\textit{Cube}, \textit{Skewed}, and \textit{BR1}). One plausible hypothesis is that the \textit{BR2} case contains a stronger shear layer and sharper velocity gradients than any of the other cases, so the ML model is forced to extrapolate and thus produces an inadequate $Pr_t$ field.

\subsection{Error metric}

The previous qualitative analysis can be complemented by a more quantitative comparison of the different temperature fields using a 3D error metric. The same error metric proposed in Milani et al. \cite{milani_interpretation2018} is used, consisting of a weighted average of the absolute difference between the LES mean temperature $\bar{\theta}_{LES}^{(i)}$ and the model predictions for the mean temperature $\bar{\theta}^{(i)}$ in each cell $i$. The volume of each cell $V^{(i)}$ is used as the weight. Formally, this metric is defined in Eq.~\ref{eq_errormetric}.
 
\begin{equation}
error = \frac{\sum_{i} \vert \bar{\theta}^{(i)} - \bar{\theta}_{LES}^{(i)} \vert V^{(i)}}{\sum_{i} V^{(i)}}
\label{eq_errormetric}
\end{equation}

\begin{table}[b]
\caption{Error metric ($\times 100$) in \textit{BR1} and \textit{BR2} cases. Only cells where $\bar{\theta}_{LES} > 0.05$ and $Y > 0$ are considered. Additionally, the Injection (\textbf{Inj}) region of interest requires $X/D < 4$ and the \textbf{Wall} region of interest requires $Y/D < 0.1$.}
\begin{center}
\label{table-errormetrics}
\begin{tabular}{| p{1.55cm} || p{0.72cm}| p{0.72cm} | p{0.72cm} || p{0.72cm}| p{0.72cm} | p{0.72cm} |}
\hline
\multirow{2}{*}{Model used} & \multicolumn{3}{c||}{\textbf{BR1}} & \multicolumn{3}{c|}{\textbf{BR2}} \\
\cline{2-7}
& \textbf{Total} & \textbf{Inj} & \textbf{Wall} & \textbf{Total} & \textbf{Inj} & \textbf{Wall} \\
\hline
$Pr_t=0.85$ & $2.96$ & $5.02$ & $6.27$ & $1.90$ & $6.17$ & $3.16$\\
\texttt{RF\symbol{95}b1} & $1.56$ & $3.33$ & $1.67$ & $4.18$ & $13.2$ & $4.92$\\
\texttt{RF\symbol{95}b2} & $2.12$ & $4.37$ & $1.56$ & $0.94$ & $2.08$ & $0.97$\\
\texttt{RF\symbol{95}s} & $2.67$ & $4.74$ & $2.68$ & $2.31$ & $9.41$ & $4.61$\\
\texttt{RF\symbol{95}c} & $7.66$ & $10.3$ & $4.46$ & $3.86$ & $11.4$ & $3.93$\\
\texttt{RF\symbol{95}csb1} & \textit{n/a} & \textit{n/a} & \textit{n/a} & $1.99$ & $7.78$ & $4.15$\\
\texttt{RF\symbol{95}csb2} & $2.48$ & $4.58$ & $2.98$ & \textit{n/a} & \textit{n/a} & \textit{n/a}\\
\hline
\end{tabular}
\end{center}
\end{table}

Table~\ref{table-errormetrics} contains the error metric calculated in \textit{BR1} and \textit{BR2} for distinct models. The summation of Eq.~\ref{eq_errormetric} is performed over all cells in three different regions of interest (Total, Injection, and Wall) which are described in the table caption. 

The values calculated for \textit{BR1} show that using a uniform $Pr_t$ does not produce good results, particularly close to the wall. All different ML models improve on it, except for the one trained only on the \textit{Cube} which produces very poor results. Most significant are the improvements in the near-wall region. However, as mentioned before, even the perfectly trained ML model (\verb|RF_b1|) cannot reduce the errors as much as one might hope, which suggests a better model form would benefit turbulent heat flux predictions in this case.

The numbers from the \textit{BR2} case also reinforce the qualitative conclusions. It is noteworthy how much reduction in error metric one can achieve by using the perfectly trained ML model, \verb|RF_b2|, suggesting the GDH is appropriate for this case. However, none of the models which are attempting to generalize perform nearly as well, and almost all seem worse than the baseline $Pr_t=0.85$ model. This shows that the training set containing \textit{Cube}, \textit{Skewed}, and \textit{BR1} does not contain all the physical information needed to learn the physics relevant to the \textit{BR2} case.

\section {VISUALIZATION}

One of the interesting conclusions from the previous sections is that a model trained on the \textit{BR2} case can generalize reasonably well to \textit{BR1}, but a model trained on the \textit{BR1} case performs very poorly on \textit{BR2}. We would like to understand when models trained on one flow will generalize to another flow. One reason why a model might fail to generalize is that it is being forced to extrapolate. If a test case contains mean flow features not found in the training set, then the model might be expected to perform poorly in such regions of extrapolation. To better understand whether the models are extrapolating, it would be useful to visualize the input parameter spaces for the \textit{BR1} and \textit{BR2} cases. To achieve this, two different dimensionality reduction techniques are employed, namely Principal Component Analysis (PCA) and t-Distributed Stochastic Neighbor Embedding (t-SNE). 

As explained in section 2.1, each computational cell in a flow configuration is seen by the machine learning model as a 19-dimensional point, in which the 19 entries form an invariant basis that depends on the mean velocity gradient, mean concentration gradient, eddy viscosity, and distance to the wall. These 19 features $\phi_i$ are then mapped to the local $Pr_t$ through a random forest. To qualitatively understand where cells coming from different flows are located in this 19-dimensional space, one common technique in the machine learning community is to apply dimensionality reduction. Each point $(\phi_1, \phi_2, ..., \phi_{19})$ is projected onto a 2-dimensional space with resulting coordinates $(\xi_1, \xi_2)$ so that they can be visualized in a plot. This operation invariably leads to information loss; different techniques try to minimize such losses in distinct ways.

Principal Component Analysis (PCA) is a widely used dimensionality reduction technique \cite{pca_tutorial}, where the lower-dimensional coordinates are constructed as a linear combination of the original coordinates. The coefficients of this linear combination are chosen such as to maximize the variance along the new axes. The principal components, which form the axes $\xi_1$ and $\xi_2$ in PCA, can be calculated efficiently using eigenvector analysis. 

On the other hand, the t-Distributed Stochastic Neighbor Embedding (t-SNE) proposed by Maaten and Hinton \cite{maaten_tsne} is a more recent technique and a very popular one in the machine learning community. It creates a non-linear mapping between the original high-dimensional space and the lower-dimensional space that tries to preserve pairwise distance between points. It is significantly more computationally expensive because gradient descent must be used to minimize the cost function. However, it typically produces plots that maintain much of the cluster-based structure of the high-dimensional space. Recently, Wu et al. \cite{wu2017_visualization} applied t-SNE visualization to high-dimensional data from turbulent flow simulations and were able to capture physically relevant differences between distinct flows.

Before applying either technique, the features were normalized to have standard deviation of 1. This is done so that distances between points in the high-dimensional space are roughly equivalent in all directions, instead of being dominated by the features that vary across the broadest range. PCA was applied to all points in the two datasets (27 million points); t-SNE was only applied in cells located on the center spanwise plane of either dataset, downsampled randomly at 5\% (totalling 27k points). The t-SNE algorithm has an important hyperparameter called perplexity, which corresponds roughly to the number of neighbors each point is expected to have \cite{maaten_tsne}. We tested different values of perplexity and found that the resulting plots are sensitive to its choice, contradicting what is suggested in Maaten and Hinton \cite{maaten_tsne}. For this type of data (subsampled points from fluid mechanics simulations), we recommend setting the perplexity to about 1-5\% of the total number of points t-SNE is applied to. In the following results, perplexity is set to 1200.

\begin{figure}[h]
\begin{center}
\includegraphics [width = 90mm]{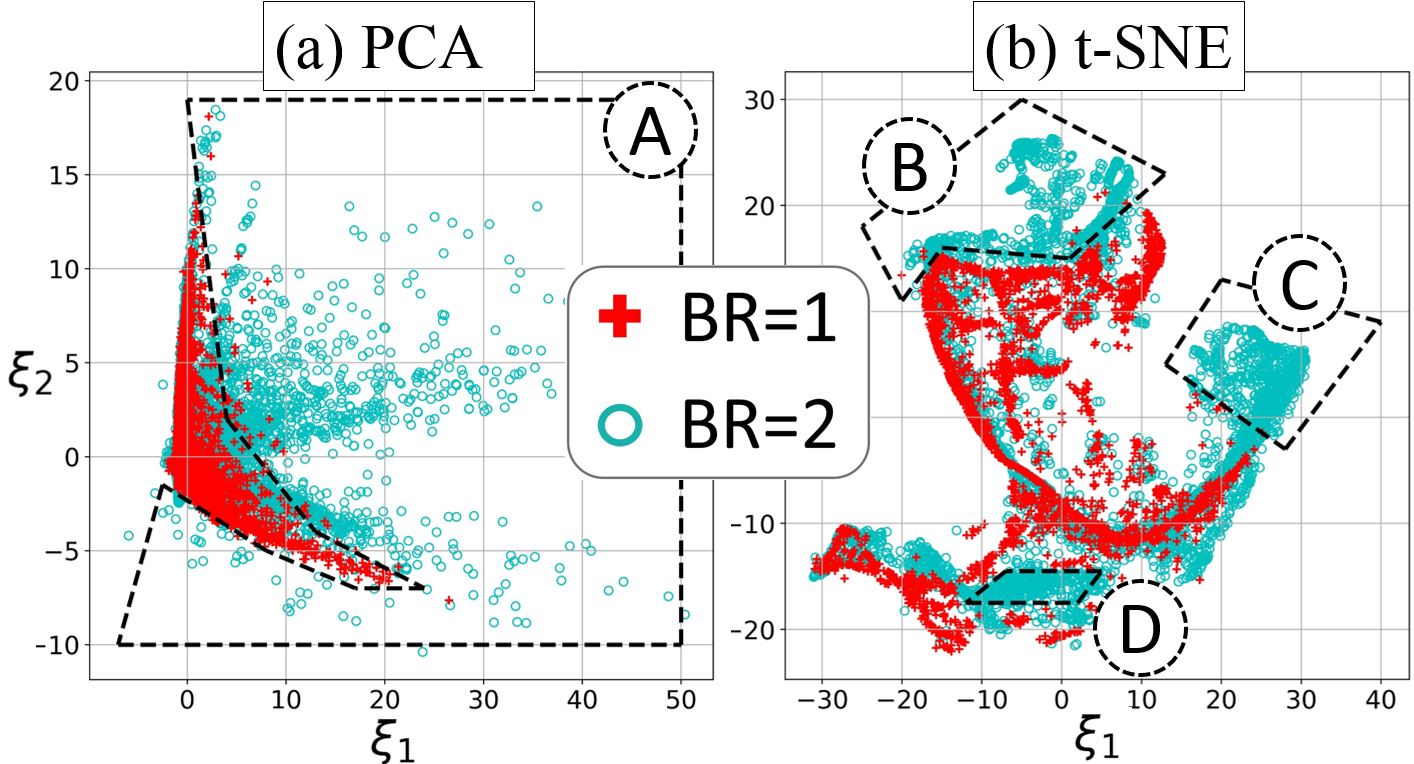}
\end{center}
\caption{Points from the \textit{BR1} and \textit{BR2} datasets mapped onto a 2-dimensional space $(\xi_1, \xi_2)$ via (a) PCA and (b) t-SNE algorithms. Four areas (A,B,C,D) are marked with dashed polygons in the two plots, consisting of regions with \textit{BR2} points but almost no points from \textit{BR1}.} 
\label{figure-dim_red} 
\end{figure}

Figure~\ref{figure-dim_red} shows points from \textit{BR1} and \textit{BR2} in a lower dimensional space, after the application of PCA (Fig.~\ref{figure-dim_red}(a)) and t-SNE (Fig.~\ref{figure-dim_red}(b)). Interestingly, both figures tell a similar story: everywhere where \textit{BR1} points are located, points from \textit{BR2} can also be found. However, there are significant regions with \textit{BR2} points but no \textit{BR1} points. This supports the hypothesis that the points of the \textit{BR1} case occupy a region which is a strict subset of the region occupied by the \textit{BR2} points in the high-dimensional space. Therefore, ML models trained on \textit{BR2} are expected to generalize well on \textit{BR1}, but the reverse is not true.

Figure~\ref{figure-dim_red} also show four specific regions of the 2D plane (A, B, C, D) that contain \textit{BR2} points but almost no points from \textit{BR1}. One of the drawbacks of PCA, which has been reported before (e.g. \cite{wu2017_visualization}), becomes clear: it tends to heavily cluster points, reducing its resolution to contrast datasets. Region A occupies most of the plot, but encompasses only about 3.5\% of all points in the \textit{BR2} case. The remaining 96.5\% seem to be located in areas well supported by the \textit{BR1} dataset. On the other hand, t-SNE tends to distribute points more evenly, and it also shows significant areas without any red points. Region B contains about 20\%, region C contains about 16.5\%, and region D contains about 8\% of all \textit{BR2} points plotted.

\begin{figure}[h]
\begin{center}
\includegraphics [width = 90mm]{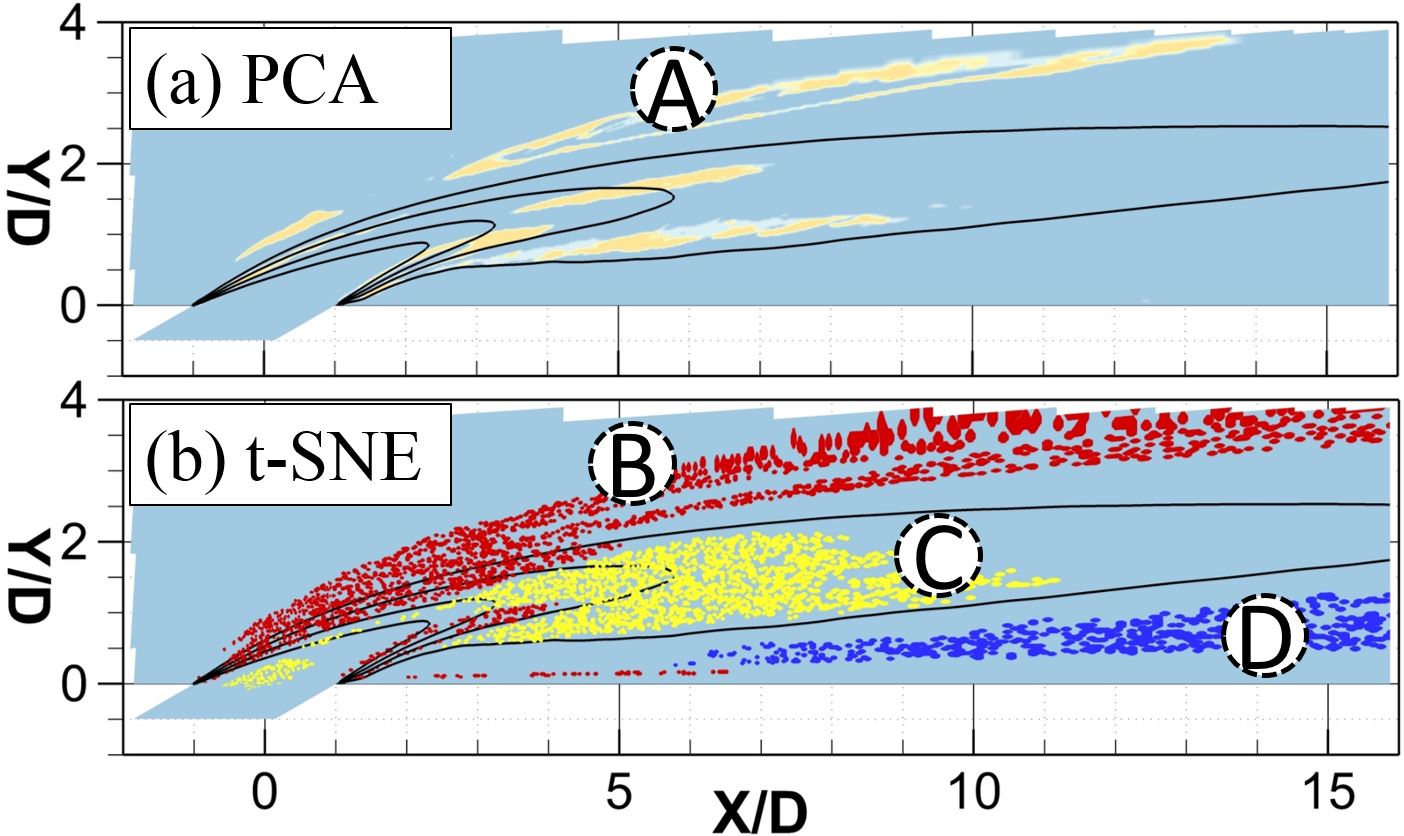}
\end{center}
\caption{Center spanwise plane of the \textit{BR2} case with black lines corresponding to isocontours of LES mean temperature at $\bar{\theta}=0.2, 0.4, 0.6, 0.8$. (a) shows points from region A (beige) in the PCA plot, and (b) shows points from regions B (red), C (yellow), and D (dark blue) in the t-SNE plot. Areas outside any of the dashed polygons in Fig.~\ref{figure-dim_red} are shown in light blue.}
\label{figure-dim_red_regions} 
\end{figure}

Each marker in Fig.~\ref{figure-dim_red} illustrates the 2D representation of the 19 features extracted from each computational cell. A natural way to investigate these plots is to determine from where in the flow the regions of extrapolation (A, B, C, and D) originate. Figure~\ref{figure-dim_red_regions} shows exactly this. It contains center spanwise planes from the \textit{BR2} dataset, and highlights the specific cells contained within regions A, B, C, and D. These are regions that seem poorly supported by the \textit{BR1} data, and thus flow locations where the random forest trained exclusively on the \textit{BR1} case is expected to generate poor results. 

Figure~\ref{figure-dim_red_regions}(a) evidences only sparse areas of the flow, mostly focused in the top shear layer and in the jet core around $X/D=5$. Comparing the top shear layer in Figs.~\ref{figure-br2_prt}(a) and \ref{figure-br2_prt}(b), it is clear that the $Pr_t$ prediction of \verb|RF_b1| is indeed very poor there. Figure~\ref{figure-dim_red_regions}(b) contains more comprehensive results, that are also consistent with the predictions shown in Fig.~\ref{figure-br2_prt}. Region B encompasses most of the shear layer on the windward side of the jet and region C highlights the jet core from $X/D=3$ to $X/D=9$. Region D shows the bottom part of the jet further downstream, starting at about $X/D=7$. In Figs.~\ref{figure-br2_prt}(a) and \ref{figure-br2_prt}(b), it is again clear that regions B and C correspond to areas where \verb|RF_b1| is particularly bad at predicting adequate $Pr_t$ values. 

Finally, it is interesting to analyze areas in Fig.~\ref{figure-dim_red_regions}(b) outside of regions B, C, and D. These include a thin region close to the bottom wall along the full domain, and the bottom half of the jet right after injection (with $Y/D < 0.5$ and $X/D < 6$). These are areas where the prediction of \verb|RF_b1| is qualitatively better, as can be seen in Fig. \ref{figure-br2_prt}(b): it correctly predicts low $Pr_t$ in the thin strip above the bottom wall, and generally higher values of $Pr_t$ right after injection when $Y/D < 0.5$. Overall, these results show that the dimensionality reduction techniques presented here are useful in identifying datasets and particular regions where generalization would and would not be expected.

\section {CONCLUSION}

In this work, a machine learning model for turbulent heat flux was presented and its ability to generalize was investigated. The framework, which is an improved version of the one presented by Milani et al. \cite{milani_approach2017}, consists of using the GDH coupled with a random forest algorithm to prescribe a non-uniform turbulent Prandtl number. The RF is trained with high-fidelity simulations and is expected to improve mean temperature results in unseen film cooling flows, which was demonstrated in this paper.

Among other findings, the results shown here suggest that the simple GDH with a perfect $Pr_t$ field improves predictions in some locations of the film cooling dataset with $BR=1$, particularly close to the wall, but it is still a deficient model in others. However, when the blowing ratio is doubled to $BR=2$, the GDH with an ideal $Pr_t$ field produced excellent results everywhere. Also, using the different datasets available here produced a model that can generalize well to the \textit{BR1} case, but did not generalize well to the higher blowing ratio case, \textit{BR2}. Two distinct dimensionality reduction techniques were used to visualize the \textit{BR1} and \textit{BR2} datasets in an attempt to explain this. The results from both of them, PCA and t-SNE, suggest that the points from the higher blowing ratio case occupy a region which is a superset of that occupied by the lower blowing ratio case in the high-dimensional feature space.

In future work, the ML approach should be used with a more advanced, anisotropic model for turbulent heat flux. The results also suggest that an effective model for film cooling must be trained with datasets spanning different blowing ratios, going at least as high as the highest blowing ratio of the configuration of interest. Finally, the visualization techniques shown in Section 5 can be leveraged to identify potentially useful new training datasets for a particular target flow.

\begin{acknowledgment}

This research was generously supported by Honeywell Aerospace. In particular, we greatly benefited from discussions with Samir Rida, Khosro Molla Hosseini, and Ardeshir Riahi.

\end{acknowledgment}


\bibliographystyle{asmems4}

\bibliography{Milani_Bib_master}

\end{document}